\documentclass[useAMS,usenatbib]{mn2e}
\usepackage{times}
\usepackage{graphicx}
\usepackage{epstopdf}
\usepackage{longtable} 
\usepackage{tablefootnote} 
\usepackage{changes}  
\usepackage{cancel}
\usepackage{ulem}
\usepackage{textcomp}
\usepackage{epstopdf}
\newcommand{\medd}{$\mathrm{\dot{M}_{Edd}}$}
\newcommand{\msun}{$\mathrm{M_{\odot}}$}
\newcommand{\msune}{$\mathrm{M_{\odot}}$}
\newcommand{\mdot}{$\mathrm{\dot{M}}$}
\newcommand{\mdote}{$\mathrm{\dot{M}}$}
\newcommand{\acef}{$\mathrm{\eta_{acc}}$}
\newcommand{\acefe}{$\mathrm{\eta_{acc}}$}

\newcommand{\mdota}{$\mathrm{\dot{M}_a}$\ }
\newcommand{\mign}{$\mathrm{\Delta M_{ign}}$}
\newcommand{\dotm}{$\mathrm{\dot{M}}$}

\def\sne{SNe~Ia}
\def\snia{SN~.Ia}

\def\am{AM~CVn}
\def\elm{ELM~WD} 
\def\apgt{\ {\raise-.5ex\hbox{$\buildrel>\over\sim$}}\ }
\def\aplt{\ {\raise-.5ex\hbox{$\buildrel<\over\sim$}}\ }
\newcommand{\ms}{\mbox {$\mathrm M_{\odot}$}}

\newcommand{\rsun}{\mbox {$\mathrm{R_{\odot}}$}}
\newcommand{\ls}{\mbox {$\mathrm{L_{\odot}}$}}
\newcommand{\rs}{\mbox {$\mathrm{R_{\odot}}$}}
\newcommand{\myr}{\mbox {~${\rm M_{\odot}~yr^{-1}}$}}
\newcommand{\porb}{\mbox {$P_{\mathrm orb}$}}

\title[He-accreting WDs]
      {He-Accreting WDs: \am\ stars with WD Donors}
\author[L. Piersanti et al.]{L. Piersanti$^{1,4}$\thanks{E-mail:
piersanti@oa-teramo.inaf.it (LP); tornambe@oa-teramo.inaf.it (AT); lry@inasan.ru (LY)} 
and L.R. Yungelson$^{2}$ and A. Tornamb\'e$^{3}$ \\
$^{1}$INAF-Osservatorio Astronomico di Collurania Teramo
              via Mentore Maggini, snc, 64100, Teramo, IT\\
$^{2}$Institute of Astronomy, Pyatnitskaya 48, 119017 Moscow, Russia\\
$^{3}$INAF-Osservatorio Astronomico  di Roma
              via di Frascati, 33, 00040, Monte Porzio Catone, IT\\
$^{4}$INFN-Sezione di Napoli, 80126 Napoli, Italy}
\begin{document}

\date{\today}

\pagerange{\pageref{firstpage}--\pageref{lastpage}} \pubyear{2013}

\maketitle

\label{firstpage}

\begin{abstract}
We study the physical and evolutionary properties of the ``WD family'' of \am\, stars 
by computing realistic models of Interacting Double-Degenerate systems. 
We evaluate self-consistently both the mass transfer rate from the donor, as determined 
by gravitational wave emission and interaction with the binary companion, and the thermal response of the accretor 
to mass deposition. 
We find that, after the onset of mass transfer, all the considered systems undergo 
a strong non-dynamical He-flash. However, due to the compactness of these systems, the expanding accretors fill their Roche 
lobe very soon, thus preventing the efficient heating of the external layers of the accreted CO WDs. 
Moreover, due to the loss of matter from the systems, the orbital separations enlarge and mass transfer comes to a halt. 
The further evolution depends on the value of \mdot\, after the donors fill again their lobe. 
On one hand, if the accretion rate, as determined by the actual value of $\mathrm{(M_{don},M_{acc})}$, is high enough, 
the accretors experience several He-flashes of decreasing strength and then quiescent He-burning sets in. Later on, 
since the mass transfer rate in IDD is a permanently decreasing function of time,
accretors experience several recurrent strong flashes. 
On the other hand, for intermediate and low values of \mdot\, the accretors enter directly 
the strong flashes accretion 
regime. As expected, in all the considered systems the last He-flash is the strongest one, even if the physical conditions 
suitable for a dynamical event are never attained. 
When the mass accretion rate decreases below $(2-3)\times 10^{-8}$\myr, the compressional heating of the He-shell 
becomes less efficient than the neutrino cooling, so that all the 
accretors in the considered systems evolve into massive degenerate objects. 
Our results suggest that \snia\, or type Ia Supernovae due to Edge-Lit Detonation in the WD family of 
\am\, stars should be much more rare than previously expected.
\end{abstract}



\section{Introduction}
\label{s:intro}
\am\, stars are ultracompact cataclysmic binaries with spectra dominated by helium. 
At the time of writing 43 confirmed and candidate objects were known, 
see Table 1 in \citet{2015MNRAS.446..391L}, and \citet{2014ATel.6669....1W,2015PASJ...67L...2K}.
Measured orbital periods range from 5.5 to 65~min. There exist also several 
cataclysmic variables (some with hydrogen-deficient spectra) below the conventional
minimum \porb\ of CV (70-80) min., which may be \am\ stars in making 
\citep{2012MNRAS.425.2548B,2013MNRAS.431..372C,2013AJ....145..145L,2014MNRAS.438..789R,2014ATel.6287....1G}.   
The significance of \am\ stars stems from their importance for the  
studies of very late stages of evolution of binary stars and of 
accretion disks physics; as well they are considered primary targets and   
verification sources for high-frequency gravitational waves detectors. 
The current models of \am\, stars envision a semidetached binary 
harbouring a carbon-oxygen white dwarf (CO WD) accreting He-rich 
matter. The donor may be either a helium WD or a low-mass helium star or a core of 
a main-sequence star strongly evolved prior to Roche-lobe overflow (RLOF).
The evolution of \am\ binaries is driven by angular momentum loss via gravitational 
waves radiation (GWR). 
An overview and a discussion of observational features, formation and evolution of these 
stars, as well as models for their disks may be found, e. g., in   
\citet{2003cvs..book.....W,nelemans_amcvn05,2009CQGra..26i4030N,2010PASP..122.1133S,2010ApJ...717.1006R,
2012A&A...544A..13K,2013GWN.....6....4A,2014LRR....17....3P,2015ApJ...803...19C}.
The topic of the present study are \am\ systems with WD donors,
sometimes also called ``interacting double-degenerates'' (IDD) or 
``white-dwarf family of \am\ stars''. 

An essential issue defining the formation of \am\ stars is stability of mass-transfer 
by degenerate donors \citep{ty79a,npv+01,mns04}. In double-degenerate systems, Roche lobe 
is first filled by the lighter of two WDs (in the context of this section -- the secondary 
with mass $M_2$). Since for WDs the power of the mass-radius relation is close to $-1/3$, 
systems with initial $q=M_2/M_1 > 2/3$ are dynamically unstable. 
The components separation $a$ of nascent  \am's is so small that mass-exchange almost 
definitely begins in the direct-impact mode, without formation of the disk. 
Strict condition of stability for the case of no feedback of the angular momentum of the 
accreted material to the orbit becomes
\begin{equation}
q <  1 +(\xi_2 - \xi_L)/2-\sqrt{(1+q) r_c},
\end{equation}
where $\xi_2$ and  $\xi_L$ are logarithmic derivatives of the radii of the donor and its
Roche-lobe with respect to its mass, $r_c \equiv R_c/a$ is the relative 
circularisation radius. In between these two limits, dynamical stability of mass-transfer 
depends on the efficiency of spin-orbit coupling. 

An additional complication is brought in by the fact that the surface luminosity of the WD, 
as determined 
by the mass deposition in the gravitational field of the accretor, can not exceed the 
Eddington limit.   
In the specific case of the gravitational potential of 
interacting double-degenerates, if \mdota is 
super-Eddington, the excess of the matter remains in the potential well of the accretor,
the released energy heats it  and 
may cause it to expand and form a common envelope in which the two components will merge \citep{hw99}. 
Numerical experiments by \citet{mns04} showed that in the case of weak tidal coupling stable mass 
transfer is possible if initial mass ratio is $\aplt 0.25$ for $M_{\rm CO} \approx 0.6$\,
\msun\, and $\aplt 0.21$\ for $M_{\rm CO} \approx 1.0$\,\ms. In the limit of very strong tidal 
coupling, the critical values of mass ratios become $\approx 0.45$ and $\approx 0.3$, respectively. 

The Eddington luminosity limit for \mdot\ is more stringent than the dynamical stability one; 
therefore  initial 
mass-transfer rates in \am\ systems should be between $10^{-6}$ and  $10^{-5}$\,\myr. However, it is 
possible that, even if \mdota $<$ \medd, it could exceed the maximum rate 
of He-burning at the base of the He-envelope \citep{nom82a}. In the latter case a red-giant like 
extended envelope forms and components, most probably, merge\footnote{We note that a 
rigorous consideration of the net angular momentum change following ballistic ejection of a particle, 
its motion and accretion \citep{2014ApJ...785..157S}, as well as the accounting for angular momentum 
exchange between the spins of the components and the orbit \citep{2015arXiv150206147K} resulted in a slight 
change of the above-mentioned conditions for dynamical stability of 
mass transfer. In any case, the effects of the Eddington luminosity and the existence of 
a limiting He-burning rate are of higher value than these corrections.}.
Thus, the existing interacting double-degenerates should start their evolution with \mdot$\sim 10^{-6}$\,\myr\ and 
may evolve in a Hubble time to \mdot$\sim 10^{-12}$\,\myr\ 
as it is theoretically inferred 
\citep{ty79a,npv+01,2007MNRAS.381..525D} and supported by the analysis of the spectra of several \am\ stars
\citep{2010AIPC.1273..305K,2014A&A...562A.132G}.

Along their path in $\mathrm{M_{acc}}$--\mdot\ plane, accretors of IDD may experience stable He-burning, burning 
via mild and strong flashes and, in principle, enter the regime of dynamical flashes
\citep[for a systematical study of the burning regimes of the accreted He see][henceforth -- Paper I]{2014MNRAS.445.3239P}. 
We define a flash ``strong'', if in its course the WD overflows its
Roche lobe. Otherwise, the flash is defined as ``mild''.
Interpolation between the models computed with constant \mdota\, suggests 
that in the course of the evolution \am\, stars experience $\sim 10$ strong non-dynamical flashes.

\citet{2007ApJ...662L..95B} paid attention to the circumstance that, with decreasing 
\mdot, the mass to be accreted to get He-ignition, \mign,  increases.
Thus, there should exist  the strongest ``last flash''. 
Further, the mass of the donor becomes smaller than  \mign.
The last flash may become dynamical and result in a detonation,
if the thermonuclear timescale, $\tau_{nuc} = c_P T/\epsilon_{nuc}$ becomes shorter than the local 
dynamical time, $\tau_{dyn} = H_P/c_s$, where $H_P$ is the pressure scale-height and $c_s$ is 
the sound speed. 
Due to the physical conditions existing in the accreting WD, the detonation 
produces short-living radioactive isotopes of Cr, Fe, and Ni. As well, ejected mass is small. 
The brightness of the event is comparable to subluminous \sne, but its rise-time is only 2-10 day.
Extrapolating the measured local birthrate of \am\ stars and assuming that all of them produce
a visible event, \citet{2007ApJ...662L..95B} estimated that the latter may occur once in 5000 -- 15000 yr 
in a $10^{11}$\,\ms\ E/S0-type galaxy., i. e., at a rate of $\sim 1/10$th of the inferred 
\sne\ rate in them. Having in mind the low luminosity of these faint supernovae and the quoted occurrence rate, 
\citeauthor{2007ApJ...662L..95B} dubbed them ``SN~.Ia'' (1/10th of brightness at 1/10th of total rate).  
According to later calculations \citep{2009ApJ...699.1365S}, such faint thermonuclear SNe are likely 
to occur if masses of accretors are in the range (0.8-1.2)\,\ms. We note that, according to existing population 
synthesis results \citep{npv+01}, \am\ stars with such masses of the accretors should be extremely rare. 
Therefore, SNe~.Ia may be substantially more rare events than estimated by Bildsten et al. 
At the moment, none of several suggested SN~.Ia candidates is definitely confirmed \citep{Drout_SN.Ia_13}.
In the model of evolution of IDD above, it was assumed that {\sl mass-transfer is continuous}. 

Recently, the authors of the present study suggested to account in the computation for the effect of 
mass and momentum loss from the binaries due to the Roche lobe overflow by the accretor during outbursts  (Paper~I).
It was assumed that the matter leaving the system has specific angular momentum of 
accretor\footnote{To some extent, a similar evolutionary scheme was 
considered for semidetached WD+He-star systems by \citet{yoon_langer04} and \citet{brooks2015}; as well, 
it resembles the scenario of evolution of cataclysmic variables with ``hibernation''.}. 

In Paper~I a grid of accretor masses and constant accretion rates was explored. In the present paper 
we consider the response to the accretion for time-dependent \mdot, appropriate to \am\ stars 
with WD-donors. As we show below, the RLOF episode and the associated mass and 
angular momentum loss from the system lead to 
the interruption of mass-transfer, thus resulting, after the flash, 
in epochs of cooling of accretor which change the character of 
thermal flashes and the course of evolution of the binaries under consideration. 
In Sect.~\ref{s:model} we present our model and justify the selection of computed evolutionary sequences; 
results are described in Sect.~\ref{s:result}. Discussion and conclusions follow in Sect.~\ref{s:diss}.

\section{Selection of initial binaries and computational assumptions}
\label{s:model}

Considerations of the mass transfer stability, of the limits imposed onto the initial \mdot\, by 
the Eddington luminosity and of the possible formation of red-giant like envelope, suggest 
that the  precursors of
WD-family of \am\ stars are detached double-degenerate (WD+WD) 
systems with rather extreme mass ratios and massive accretors. 
Such binaries are observed as detached binary ``extremely low-mass'' white dwarfs (ELM),
which harbour a He WD with mass below $\simeq 0.2$\,\ms\ and a much more massive (presumably) CO companion;
some of them have merger time less than the Hubble time 
\citep[see, e. g.,][]{2013ApJ...769...66B,2014MNRAS.438L..26K}. 

Currently, several systems are observed, which may be considered as proper candidate 
precursors of \am\, stars (note, here
we call ``primary'' the more massive invisible component with $M_1$).
The first candidate is eclipsing binary  SDSS~J075141.18-014120.9 
with \porb = 1.9 hr, $(M_2 + M_1)/\ms = (0.19 \pm 0.02) + (0.97^{+0.06}_{-0.01}$)  and expected 
merger time  $\simeq$160~Myr \citep{2014MNRAS.438L..26K}.
Another ``best'' candidate is the eclipsing detached binary NLTT~11748 with 
\porb = 5.6\,hr  and estimated mass of the secondary ranging from  
$M_2 = (0.136 \pm 0.007)$ to $(0.162 \pm 0.007)$\,\ms\ and, correspondingly, 
mass of the primary from $M_1 = (0.707 \pm 0.008)$ to $(0.740 \pm 0.008)$\ms\ 
\citep{2014ApJ...780..167K}.
Several candidate systems are single-lined and for them only lower limits of $M_1$ are available. 
Among them are 
SDSS~J1741+6526 -- \porb = 1.47~hr, $M_2 = 0.17$\,\ms, $M_1 \geq 1.11$\,\ms\ 
\citep{2014MNRAS.438L..26K};
SDSS~J1141+3850 --  $\porb \approx 6.23$~hr, $M_2 = 0.17$\,\ms, $M_1 \geq 0.76$\,\ms;
SDSS~J1238+1946 -- $\porb \approx 5.46$~hr, $M_2 = 0.17$\,\ms, $M_1 \geq 0.64$\ms; 
SDSS~J2132+0754 --  $\porb \approx 6.01$~hr,  $M_2 = 0.17$\,\ms, $M_1 \geq 0.95$\,\ms\ 
\citep{2013ApJ...769...66B}. Yet another candidate with less certainly estimated parameters is
SDSS~1257+5428 --  \porb = 4.55~hr, $M_2\sim 0.2$\,\ms, $M_1 \sim 1$\,\ms\ 
\citep{2010ApJ...719.1123K,2011ApJ...736...95M}.
At the end, other candidate \am\ systems may be hidden among observed sdB stars, deemed to evolve into
CO WDs with mass $\apgt 0.5$\,\msun, with low-mass WD companions, like 
PG~1043+760 -- $\porb=2.88$~hr, $M_2 \geq 0.101$,\ms,
SDSS~J083006+47510 -- $\porb=3.552$~hr, $M_2 \geq 0.137$,\ms\ \citep{2015A&A...576A..44K}. 

Having in mind the parameters of candidate \am\ systems and for the sake of comparison with computations
in Paper~I, in the current work we selected  the following 
donor and accretor combinations:
$\mathrm{(M_{don} , M_{acc})}$=(0.17,0.60), (0.15,0.92), (0.20,1.02) \msune.
In the following, we address the sequences of models for these systems 
as S060+017, S092+015, and S102+020, respectively.
The initial CO WD models have been obtained by evolving the ``{\it heated Models}'' 
M60, M092 and M102 from Paper~I along the cooling sequence down to the luminosity level L=0.01 \ls.

\begin{table}  
\caption{Physical properties of the initial binary systems considered in the present work 
         We list the separation $a$, 
         the mass of the donor $\mathrm{M_{don}}$, the mass of the accretor $\mathrm{M_{acc}}$
         and some physical properties of the accretors, namely
         the mass extension of the He-deprived core $\mathrm{M_{CO}}$, 
         the mass extension of the more external layer where the helium 
         abundance by mass fraction is larger than 0.05 $\mathrm{\Delta M_{He}}$,
         the temperature $\mathrm{T_c}$ in K and the density $\mathrm{\rho_c}$ in 
         $\mathrm{g\,cm^{-3}}$ at the center, the surface luminosity $\mathrm{L}$, 
         the effective temperature $\mathrm{T_{eff}}$ in K and the 
         surface radius $\mathrm{R}$. In the last row we report the cooling time 
         $\mathrm{t_{cool}}$ defined as the time elapsed from the bluest point 
         along the loop in the HR diagram and the epoch of the initial model 
         considered in the current work.}
  \label{tab1}
\centering 
  \begin{tabular}{l r r r }
   \hline\hline
   LABEL                         & S060+017  & S092+015 & S102+020 \\
   $a$ (in $10^{-2}$\rsun)& 7.599 &  9.222 &  7.965 \\
   $\mathrm{M_{don}}$ (in \msune)& 0.170 &  0.150 &  0.200 \\
   $\mathrm{M_{acc}}$ (in \msune)& 0.59678 &  0.91962 &  1.020467 \\
   $\mathrm{M_{CO}}$ (in \msune)& 0.5158  & 0.9114  & 1.0158  \\
   $\mathrm{\Delta M_{He}}$ (in $10^{-2}$\msune) &  1.551 &  0.389 &  0.056 \\
   $\mathrm{\log(T_c)}$      &  7.2820 &  7.1478 &  7.2329 \\
   $\mathrm{\log(\rho_c)}$   &  6.5626 &  7.3378 &  7.6109 \\
   $\mathrm{\log(L/L_\odot)}$& -2.0077 & -2.1163 & -2.0322 \\
   $\mathrm{\log(T_{eff})}$  &  4.2102 &  4.2640 &  4.3148 \\
   $\mathrm{\log(R/R_\odot)}$& -1.9012 & -2.0633 & -2.1228 \\
   $\mathrm{t_{cool} (in 10^8 yr)}$& 1.647 & 2.767 & 2.889 \\
   \hline                  
  \end{tabular}
\end{table}  

Some relevant physical quantities for these binary WDs are listed 
in Table~\ref{tab1}. The donors are modeled as zero-temperature objects 
with Eggleton's M-R relation \citep[see][]{vr88}. In the computation we neglect possible effects of finite initial 
entropy of the He WDs \citep{del_bild_fin03,2007MNRAS.381..525D}, but we discuss them in \S4..
We do not consider the effects
related to the possible presence of a thin hydrogen layer at the surface of the 
white dwarfs when they come into contact 
\citep{2006ApJ...653.1429D,2007MNRAS.381..525D,2012ApJ...758...64K,2013ApJ...770L..35S,2015arXiv150205052S} and 
consider accretion of He only.
\begin{figure}       
 \centering
  \includegraphics[width=\columnwidth]{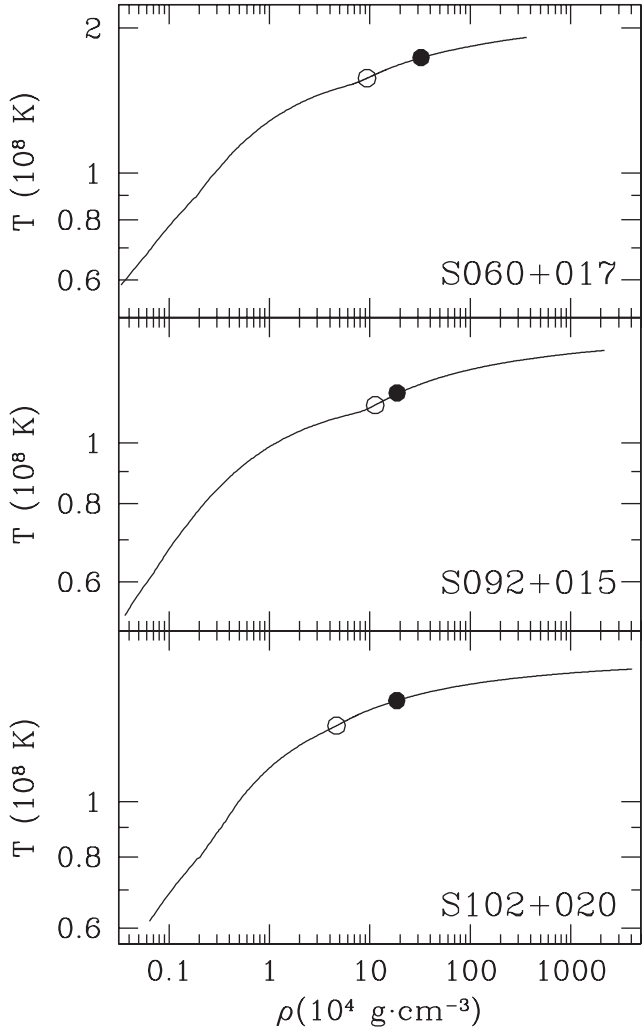}
  \caption{Profiles in the $\rho - T$ plane for the accreting CO WDs at the 
           beginning of the mass transfer process. Each panel refers to a different 
           initial binary system, as labelled. Filled circles mark the He/CO interface 
           while open ones the point where X(He)=0.05, roughly corresponding to the He-burning shell.
}
\label{fig01}
\end{figure}

\section{Results}  
\label{s:result}

At the beginning of the computation, the two components of each system
are put in contact (i. e. $\mathrm{R_{don}^{Roche}=R_{don}}$). At each time step, the angular momentum 
loss via GWR is computed, so that the donor overfills its own Roche lobe. As 
a consequence, mass is removed in order to restore the condition $\mathrm{R_{don}=R_{don}^{Roche}}$ 
and it is transferred conservatively to the accretor. If and when the accretor overfills its 
own Roche lobe, due to evolutionary reasons (e. g. the onset of a He-flash), mass is removed 
from the accretor, assuming that it is ejected from the system. 
The specific angular momentum of the lost matter is assumed to be equal to the orbital one 
of the accretor. 
The thermal response of the accretor to mass deposition is computed in detail by using an updated 
version of the FRANEC code, the original one being described in \cite{chieffi1989}. 
The setup of the code 
as well as the input physics are the same as in Paper~I. The chemical composition of the 
matter transferred from the donor to the accretor is fixed as in Paper I, 
by assuming that all CNO elements have been converted into $^{14}\mathrm{N}$, namely: 
$\mathrm{Y^{i}_{^{12}C}+Y^{i}_{^{13}C}+Y^{i}_{^{14}N}+Y^{i}_{^{15}N}+Y^{i}_{^{16}O}+Y^{i}_{^{17}O}+Y^{i}_{^{18}O}=Y^{f}_{^{14}N}}$,
where $\mathrm{Y_{j}}$ is the abundance by number of the $j$-isotope 
and the superscripts $i$ and $f$ refer to the initial MS star and the final He-donor WD, respectively.
In Fig. \ref{fig01} we show the profiles in the $\rho-T$ plane for the accretors in our models 
at the beginning of the mass transfer process {while in Fig. \ref{fig02} we plot as a function of the WD
mass fraction the mass fraction abundances of $\rm{^{4}He},\ {^{12}C}$ and $\rm{^{16}O}$ in the most 
external layers of the accretors for the same structures. 
\begin{figure}         
 \centering
  \includegraphics[width=\columnwidth]{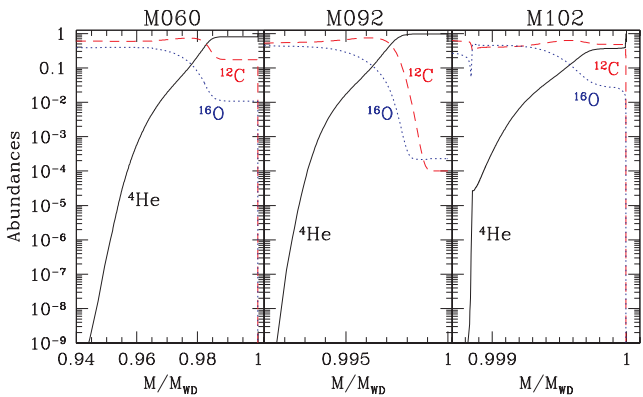}
  \caption{Mass fraction abundances of $\rm{^{4}He}$ (solid lines), $\rm{^{12}C}$ (dashed lines), and 
  $\rm{^{16}O}$ (dotted lines) in the most external layers of the accretors in the three considered 
  binary systems. Abscissa represents the mass fraction.
} 
\label{fig02}
\end{figure}

When the mass transfer starts, the evolution of the accretors in the three considered 
systems is quite similar, as illustrated in Fig. \ref{fig03}, where we show the evolutionary 
tracks in the HR diagram for the CO WDs in the investigated systems. Along the tracks, we mark 
with different symbols several important epochs. 
For each point we list in Table~\ref{tab2} some relevant physical 
properties of the accretors and of the binaries.

When the matter falls onto the accretor, it delivers gravitational energy that is locally stored as 
thermal energy and triggers the evolution backward along the cooling sequence. 
The heating determined by mass deposition drives to He-ignition\footnote{Consistently 
with Paper~I, we individuate the ignition epoch as the one when the nuclear energy per unit of time 
delivered by He-burning is 100 times the surface luminosity.}. 
The maximum 
luminosity attained during this phase depends mainly on the relative efficiency
of the heating by accretion and the inward thermal diffusion; the former is determined by the 
accretion rate, while the latter depends on the thermal structure of the He-buffer. 
As the He-burning shell has partially degenerate physical conditions, a thermonuclear runaway 
occurs, driving to a very powerful non-dynamical He-flash. 
As discussed in Paper~I, this first flash represents a sort of ``heating'' mechanism which alters 
the thermal content of the whole He-buffer above the CO core and sets in the physical 
conditions suitable for quiescent He-burning. The latter could occur only after the flashing structures 
have attained their locations in the HR-diagram corresponding to a post AGB-star with the same 
CO core and He-buffer masses. So this first He-flash represents the analog in real binary systems of the 
``heating procedure'' adopted in Paper~I. 
\begin{figure}        
 \centering
  \includegraphics[width=\columnwidth]{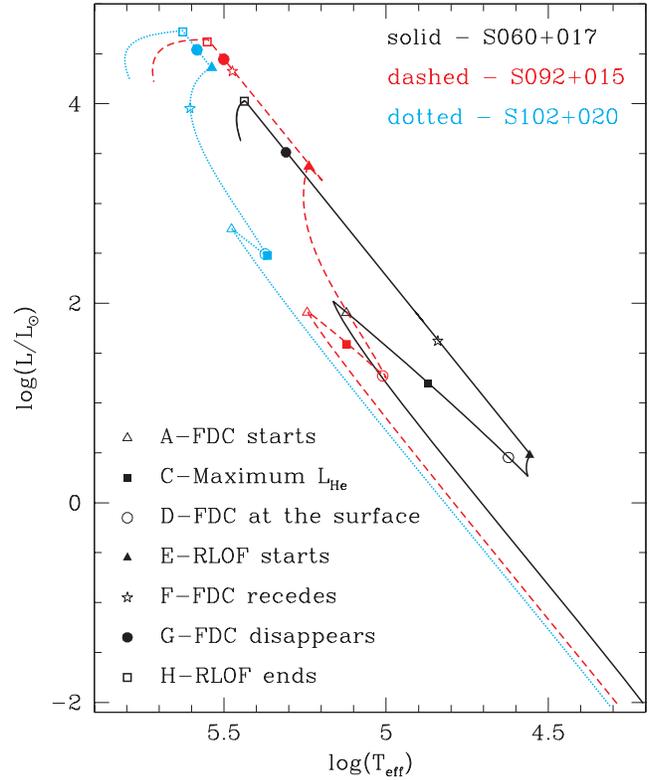}
  \caption{Evolution in the HR diagram of the accretors in the three 
  binary systems listed in Table~\ref{tab1}. Along the  tracks some points corresponding to 
  certain epochs in evolution are marked by different symbols and letters, see 
Table~\ref{tab2}.``FDC'' means flash-driven convection.
} 
\label{fig03}
\end{figure}

\begin{table*}  
\caption{Physical properties of the accretors in the binary systems listed in Table~\ref{tab1} 
         at same selected epochs during the first He-flash episode. The various times are:
         A - onset of the flash-driven convective shell; 
         B - ignition of He-burning;
         C - maximum luminosity of the He-burning shell;
         D - flash-driven convective zone attains the stellar surface;
         E - beginning of the RLOF episode; 
         F - flash-driven convective shell recedes from the stellar surface;
         G - flash-driven convective shell disappears;
         H - end of the RLOF episode; 
         I - resumption of the mass transfer from the donor.
         For each epoch of each binary system 
         we list surface luminosity $\mathrm{L/L_\odot}$ and temperature 
         $\mathrm{T_{eff}}$, the mass coordinate of the He-burning shell  
         $\mathrm{M_{He}}$ in \msune, the temperature $\mathrm{T_{He}}$\ in $10^8$ K, the density 
         $\mathrm{\rho_{He}}$\ in $\mathrm{10^4 g\cdot cm^{-3}}$ and luminosity $\mathrm{L_{He}}$ in 
         $\mathrm{erg\cdot s^{-1}}$ 
         of the He-burning shell, the separation of the system $a$ in $10^{-2}$\rs, the total mass of the 
         accretor $\mathrm{M_{acc}}$ in \msune, and of the donor $\mathrm{M_{don}}$ in \msune,
         the time elapsed between two successive epochs  $\mathrm{\Delta t}$ in yr. 
         For each considered binary system we report also the amount of mass transferred from the
         donor to the accretor $\mathrm{\Delta M_{tran}}$ in $10^{-3}$\ms, the mass lost during the RLOF\
         $\mathrm{\Delta M_{lost}}$\ in $10^{-3}$\,\ms\ and the corresponding value of the retention 
         efficiency \acefe. Note, negative $\mathrm{\Delta t}$ for events D and F in  S102+020 sequence 
         mean that they happen \textit{before} event E. The last block of data refers to the ``Cold'' S060+017 
         model described in Sect. \ref{s:cold}.}
\label{tab2} 
\centering 
  \begin{tabular}{l r r r r r r r r r}
   \hline\hline
{\it } & A & B & C & D & E & F & G & H & I \\
   \hline
  \multicolumn{10}{c}{S060+017}\\
$\mathrm{\log(T_{eff})}$          &  5.128  &  5.127  &  4.873  &  4.625  &  4.560  &  4.843  &  5.311  &  5.439  &  4.864  \\
$\mathrm{\log(L/L_\odot)}$        &  1.896  &  1.893  &  1.199  &  0.454  &  0.478  &  1.622  &  3.513  &  4.024  &  0.715  \\
$\mathrm{M_{He}/M_\odot}$         & 0.60176 & 0.60074 & 0.59945 & 0.59944 &	0.59942 & 0.59920 & 0.59801 & 0.59748 & 0.58997\\
$\mathrm{\rho_{He}}$              &  7.878  &  7.896  &  2.128  & 1.250   & 0.770   &  0.174  &  0.172  &  0.255  &  5.249  \\
$\mathrm{T_{He}}$                 &  1.235  &  1.343  &  3.260  & 3.539   & 3.579   &  2.893  &  2.183  &  1.638  &  0.605  \\
$\mathrm{\log(L_{He}/L_\odot)}$   &  2.984  &  3.894  &  10.278 & 9.745   & 9.210   &  6.406  &  4.794  &  2.738  & -6.119  \\
$a\mathrm{\ (in\ 10^{-2}R_\odot)}$&  8.592  &  8.592  &  8.593  & 8.593   & 8.593   &  8.734  & 8.8230  &  8.886  &  8.504  \\
$\mathrm{M_{acc}/M_\odot}$        & 0.62559 & 0.62560 & 0.62561 & 0.62561 & 0.62560 & 0.61316 & 0.60557 & 0.60018 & 0.60018 \\
$\mathrm{M_{don}/M_\odot}$        & 0.14119 & 0.14118 & 0.14117 & 0.14117 & 0.14117 & 0.14117 & 0.14117 & 0.14117 & 0.14117 \\
$\mathrm{\Delta t}$ (yr)          & 55637.0 & 42.4986 & 12.9458 & 0.0006  & 0.0018  & 0.3910  & 15.2389 & 54.1418 & 23945.8 \\
\hline
\multicolumn{1}{c}{\it } & \multicolumn{3}{c}{$\mathrm{\Delta M_{tran}=28.824}$} & 
\multicolumn{3}{c}{$\mathrm{\Delta M_{lost}=25.427}$} & \multicolumn{2}{c}{\acefe=0.118} \\
   \hline
  \multicolumn{10}{c}{S092+015}\\                                                     
$\mathrm{\log(T_{eff})}$          &  5.246  &  5.245  &  5.124  &  5.013  &  5.245  &  5.481  &  5.509  &  5.554  &  5.097  \\
$\mathrm{\log(L/L_\odot)}$        &  1.909  &  1.908  &  1.588  &  1.272  &  3.376  &  4.324  &  4.439  &  4.620  &  1.286  \\
$\mathrm{M_{He}/M_\odot}$         & 0.92281 & 0.92249 & 0.92205 & 0.92204 & 0.92198 & 0.92156 & 0.91634 & 0.91600 & 0.91600 \\
$\mathrm{\rho_{He}}$              & 11.594  & 11.659  &  4.156  &  3.702  &  0.724  &  0.164  &  0.171  &  0.300  &  7.417  \\
$\mathrm{T_{He}}$                 &  1.248  &  1.354  &  4.073  &  3.765  &  3.949  &  2.930  &  29098  &  2.812  &  1.062  \\
$\mathrm{\log(L_{He}/L_\odot)}$   &  3.052  &  3.910  & 11.311  & 11.177  &  8.767  &  5.462  &  5.319  &  3.910  & -4.224  \\
$\mathrm{A\ (in\ 10^{-2}R_\odot)}$&  9.651  &  9.651  &  9.651  &  9.651  &  9.651  &  9.713  &  9.718  &  9.767  &  9.615  \\
$\mathrm{M_{acc}/M_\odot}$        & 0.92985 & 0.92985 & 0.92985 & 0.92985 & 0.92985 & 0.92307 & 0.92253 & 0.91707 & 0.91707 \\
$\mathrm{M_{don}/M_\odot}$        & 0.13977 & 0.13977 & 0.13977 & 0.13977 & 0.13977 & 0.13977 & 0.13977 & 0.13977 & 0.13977 \\
$\mathrm{\Delta t}$ (yr)          & 24468.8 & 11.4436 & 3.37408 & 0.00006 & 0.00325 & 0.83678 & 3.34515 & 45.7288 & 6028.94 \\
\hline
\multicolumn{1}{c}{\it } & \multicolumn{3}{c}{$\mathrm{\Delta M_{tran}=10.230}$} & 
\multicolumn{3}{c}{$\mathrm{\Delta M_{lost}=12.787}$} & \multicolumn{2}{c}{\acefe=-0.250} \\
   \hline                  
  \multicolumn{10}{c}{S102+020}\\
$\mathrm{\log(T_{eff})}$          &  5.479  &  5.479  &  5.368  &  5.375  &  5.540  &  5.607  &  5.585  &  5.629  &  5.495  \\
$\mathrm{\log(L/L_\odot)}$        &  2.744  &  2.743  &  2.477  &  2.493  &  4.361  &  3.954  &  4.541  &  4.721  &  2.808  \\
$\mathrm{M_{He}/M_\odot}$         & 1.02090 & 1.02078 & 1.02068 & 1.02070 & 1.02051 & 1.02070 & 1.02010 & 1.02036 & 1.01983 \\
$\mathrm{\rho_{He}}$              &  4.560  &  4.535  &  1.395  &  1.445  &  0.455  &  0.716  &  0.468  &  0.133  &  2.754  \\
$\mathrm{T_{He}}$                 &  1.488  &  1.637  &  4.130  &  4.081  &  3.543  &  3.976  &  3.249  &  2.599  &  1.612  \\
$\mathrm{\log(L_{He}/L_\odot)}$   &  3.867  &  4.744  &  9.573  &  9.566  &  7.175  &  8.588  &  5.178  &  4.396  & -0.088  \\
$\mathrm{A\ (in\ 10^{-2}R_\odot)}$&  8.020  &  8.020  &  8.020  &  8.020  &  8.020  &  8.020  &  8.029  &  8.030  &  8.016  \\
$\mathrm{M_{acc}/M_\odot}$        & 1.02250 & 1.02250 & 1.02250 & 1.02250 & 1.02250 & 1.02250 & 1.02100 & 1.02067 & 1.02067 \\
$\mathrm{M_{don}/M_\odot}$        & 0.19797 & 0.19797 & 0.19797 & 0.19797 & 0.19797 & 0.19797 & 0.19797 & 0.19797 & 0.19797 \\
$\mathrm{\Delta t}$ (yr)          & 939.911 & 0.81022 & 0.26848 &-0.00001 & 0.05024 &-0.04771 & 2.90204 & 13.4218 & 181.015 \\
\hline
\multicolumn{1}{c}{\it } & \multicolumn{3}{c}{$\mathrm{\Delta M_{tran}=2.034}$} & 
\multicolumn{3}{c}{$\mathrm{\Delta M_{lost}=1.829}$} & \multicolumn{2}{c}{\acefe=0.101} \\
   \hline
  \multicolumn{10}{c}{S060+017 {\it Cold}}\\
$\mathrm{\log(T_{eff})}$          &  5.128  &  5.127  &  4.876  &  4.626  &  4.573  &  4.388  &  4.972  &  5.440  &  4.850  \\
$\mathrm{\log(L/L_\odot)}$        &  1.883  &  1.880  &  1.195  &  0.450  &  0.534  &  0.660  &  3.399  &  4.027  &  0.642  \\
$\mathrm{M_{He}/M_\odot}$         & 0.60318 & 0.60230 & 0.60089 & 0.60088 &	0.60087 & 0.60061 & 0.59746 & 0.59540 &0.59050\\
$\mathrm{\rho_{He}}$              &  7.907  &  7.883  &  2.154  & 1.258   & 0.757   &  0.186  &  0.169  &  0.266  &  4.113  \\
$\mathrm{T_{He}}$                 &  1.233  &  1.341  &  3.271  & 3.558   & 3.596   &  3.022  &  2.304  &  1.664  &  0.568  \\
$\mathrm{\log(L_{He}/L_\odot)}$   &  2.964  &  3.882  &  10.317 & 9.770   & 9.204   &  6.714  &  5.605  &  2.716  & -7.129  \\
$\mathrm{A\ (in\ 10^{-2}R_\odot)}$&  8.626  &  8.627  &  8.627  & 8.627   & 8.627   &  8.667  &  8.812  &  8.958  &  8.520  \\
$\mathrm{M_{acc}/M_\odot}$        & 0.62646 & 0.62647 & 0.62647 & 0.62647 & 0.62647 & 0.62297 & 0.61032 & 0.59803 & 0.59803 \\
$\mathrm{M_{don}/M_\odot}$        & 0.14032 & 0.14031 & 0.14031 & 0.14031 & 0.14031 & 0.14031 & 0.14031 & 0.14031 & 0.14031 \\
$\mathrm{\Delta t}$ (yr)          & 58331.2 & 45.0383 & 11.9684 & 0.0005  & 0.0019  & 0.2854  & 30.3768 & 66.2967 & 86893.4 \\
\hline
\multicolumn{1}{c}{\it } & \multicolumn{3}{c}{$\mathrm{\Delta M_{tran}=29.694}$} & 
\multicolumn{3}{c}{$\mathrm{\Delta M_{lost}=28.444}$} & \multicolumn{2}{c}{\acefe=0.042} \\
\hline                  
  \end{tabular}
\end{table*}
The energy delivered by the He-flash can not be removed via radiative transfer and, hence, 
a convective shell forms very soon (point~A in the HR diagram) and rapidly 
attains the stellar surface (point D). Hence, the huge amount of 
nuclear energy delivered by the He-flash is injected into the whole He-rich buffer above the He-burning shell, 
whose thermal content becomes too large for a compact configuration as that of the flashing objects. 
In order to dissipate the too large thermal energy stored in the He-rich mantle the accretors start to expand. 
If the flashing structure would be isolated in the space, it could evolve freely, increasing its luminosity and, then, 
should expand to very large dimensions, thus dissipating 
a part of its thermal energy via mechanical work and finally reaching 
at the He-shell the physical conditions suitable for quiescent He-burning. 
But the objects considered in the present work are embedded in compact
binary systems, so that they overfill 
their own Roche lobe very soon (point  E). The following evolution of the accretors occurs 
inside ``the Roche lobe finite space'', 
so that all the matter passing through the critical Roche surface is lost by the stars and, hence, by the binary systems. 
At the onset of the RLOF episode, due to the loss of matter and angular momentum
from the systems, the orbital separation increases, therefore,
the donors detach from their Roche lobes and, hence, the mass transfer halts. 
During the RLOF episode the accretors 
evolve at almost constant radii, increasing their surface luminosity and effective temperature 
up to when they recede definitively from their Roche lobe (point H). 
Since the components of the binary are now detached, 
so that mass transfer from donors to accretors can 
resume only after angular momentum loss by GWR shrinks the orbits, thus forcing  the donors to 
overfill once again their Roche lobe. 
During the time between the end of the RLOF and the re-onset of mass accretion, accretors 
evolve first up to the bluest point along the loop in the HR diagram and, then, down along the 
cooling sequence. 

For each considered system, in Table~\ref{tab2} we report the total mass transferred from 
the donor to the accretor  during the first mass transfer episode ($\mathrm{\Delta M_{tran}}$), 
the mass lost during the RLOF phase ($\mathrm{\Delta M_{lost}}$)
and the corresponding accumulation efficiency \acef, defined as in Eq. (5) in Paper~I: 
\begin{equation}
\mathrm{
\eta_{acc}=1-{\frac{\Delta M_{lost}}{\Delta M_{tr1}+\Delta M_{tr2}}},
}
\end{equation}
where $\mathrm{\Delta M_{tr1}}$ is the amount of mass transferred up to 
the onset of the RLOF episode and 
$\mathrm{\Delta M_{tr2}}$ the one accreted after the end of the RLOF up to the bluest point along the loop 
in the HR diagram.

\subsection{S060+017 System}
\label{s:s060}

In Fig.~\ref{fig04} we plot for the S060+017 and S092+015 systems (the latter to be 
discussed in the next subsection) the dependence of the mass transfer rate from the donor 
(upper panel) and the dependence of the slope of the curves 
$n = \mathrm{d}\log{\dot{M}}/ \mathrm{d}\log{P_{\mathrm orb}}$ (lower panel) on the orbital 
period \porb. For the sake of comparison, in the lower panel we show also the $n$ curve for 
a system with initial parameters equal to the ones of S060+017, but evolving completely 
conservatively. The Figure clearly shows that thermonuclear outbursts, interrupting the mass 
transfer (see previous Section), virtually do not influence the common 
behaviour of the \dotm - \porb\ relation for ultracompact binaries evolving under the influence
of angular momentum loss by GWR. Recently, \citet{2015ApJ...803...19C}, using the limit cycle 
accretion disk instability model, attempted to constrain semi-analytically the mass transfer rate 
in \am-stars by fitting \mdot\ to the \porb~ limits of the accretion disk-instability range of these 
systems, namely 20 and 44~min. The metallicity of \am-stars is unknown. For the range of $Z$ 
from 0.0 to 0.04, \citeauthor{2015ApJ...803...19C} found values of $n$ from $-5.38\pm0.3$ to 
$-5.06\pm0.3$. Though the period range of interest for the present work is below 20~min., the 
comparison to the results of \citet{2015ApJ...803...19C}, as well as to earlier evaluations 
of, e. g., \citet{ty96,npv+01} validates our computations. As it concerns the estimates of 
accretion rates in \am-stars, they are obtained by indirect methods and hardly can constrain
the \dotm-\porb\ relation. They may be considered as ``indicators'' of the proper order of 
magnitude of the computed \mdot, at best.
\begin{figure}   
 \centering
  \includegraphics[width=\columnwidth]{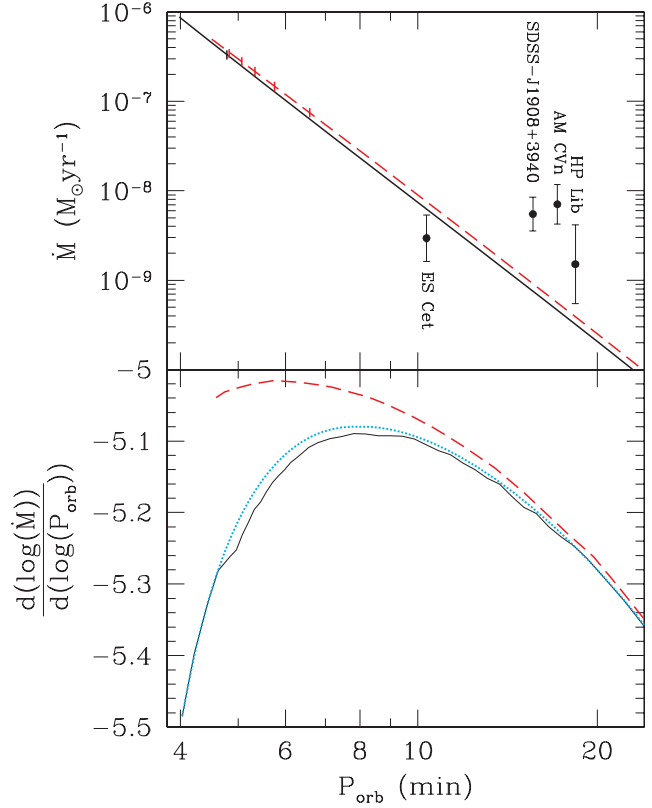}
  \caption{Upper panel: dependence on the orbital period of the mass transfer rate from 
  the donors in the systems S060+017 (solid line) and S092+015 (dashed line). The small vertical 
  ticks on the curves mark the values of \porb~ when mass transfer does not occur. In these 
  systems the first outbursts happen at \porb=4.792~min. and 4.830 min., respectively and the 
  corresponding gaps $\Delta P_{orb}$ (0.13 and 0.09 s, respectively) are too short to be 
  noticeable in the scale of the plot. Filled circles with errorbars show estimates of mass 
  transfer rates in observed \am\ systems (taken from the the compilation by 
  \citet{2012A&A...544A..13K}). Lower panel: the value of 
  $n = \mathrm{d}\log{\dot{M}} / \mathrm{d}\log{P_{\mathrm orb}}$ for the S060+017 
  and S092+015 systems (solid and dashed lines, respectively). The dotted line shows 
  the $n$ profile for the case of completely conservative mass exchange in a 
  (0.60+0.17)\,\ms\ system.
} 
\label{fig04}
\end{figure}  

In Fig. \ref{fig05} we plot the mass transfer rate (upper panel) for the S060+017 system  
as well as the evolution of the temperature (middle panel) and density (lower panel) of the 
He-shell in the accretor as a function of time. 
As it can be seen, after the He-flash and the related RLOF episode, mass transfer is not active and  
the He-burning shell cools down. When mass transfer resumes, the corresponding 
accretion rate is about $3.23\times 10^{-7}$\myr. 
Due to the deposition of matter, the He-shell starts to heat up once again, but when \mdot\, becomes 
lower than $8\times 10^{-8}$\myr, the mass deposition becomes unable to balance the cooling and the 
accretor further evolves along its cooling sequence. 
Thus, despite quite substantial amount of matter may still be transferred to the CO 
WD in this system (about 0.14\,\ms), the nuclear evolution of the binary 
is terminated. Though, it will retain its status of \am\ star.
\begin{figure}   
 \centering
  \includegraphics[width=\columnwidth]{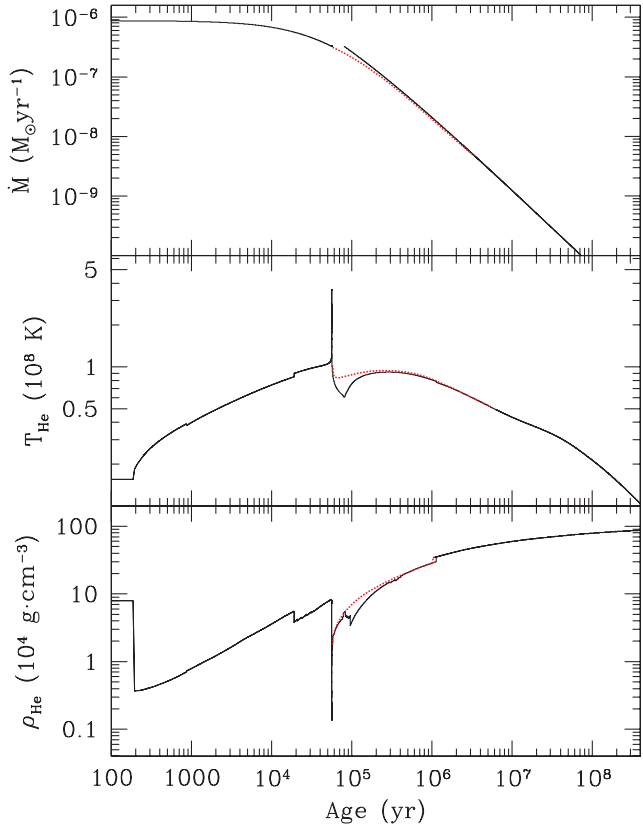}
  \caption{Time evolution of the accretion rate ($\mathrm{\dot{M}}$ - upper panel), 
  of the He-shell temperature ($\mathrm{T_{He}}$ - middle panel) and density 
  ($\mathrm{\rho_{He}}$ - lower panel) of the accretor in the binary system 
  S060+017. Dotted lines refer to the evolution of the same object, but starting to accrete 
  mass soon after the end of the RLOF. For more details see text.} 
\label{fig05}
\end{figure}

At variance with the results discussed above, the ``heated'' M060 model in Paper~I, corresponding to 
the accretor in the S060+017 system, after the first He-flash, burns helium quiescently for 
\mdote$\ge 1.5\times 10^{-7}$\myr.
However, it has been noted that the present computation differs from those in Paper~I in many regards. 
First of all, in  Paper~I the accretion rate was kept strictly constant in each simulation and mass deposition 
was restarted soon after the end of the RLOF, thus preventing substantial cooling of the He-buffer. 
In addition, the adopted value for the Roche lobe radius (10 \rsun) was 
definitively larger than the one proper to the self-consistent computation of the S060+017 system. 

In order to illustrate the origin of such a
difference, we computed two  models, arbitrarily varying 
the value of the separation during the evolution of the S060+017 system. In the first one we 
reduced the value of the separation after the end of the RLOF episode in such a way that mass 
transfer resumed immediately. The time evolution of the mass transfer rate and of the 
temperature and density of the He-burning shell for this model is displayed 
by dotted lines in Fig. \ref{fig05}. The  prompt onset of mass 
accretion prevents the decrease of the temperature of the He-burning shell below $\sim 8.3\times 10^7$ K, 
but, in any case, as \mdot\, continuously decreases, the accretor never attains the physical conditions 
suitable for the re-ignition of helium. 
This numerical experiment clearly suggests that the 
origin of the difference between the present computation and the corresponding 
one in Paper~I depends on
the different thermal content of the post-RLOF structure. 

\begin{figure}        
 \centering
  \includegraphics[width=\columnwidth]{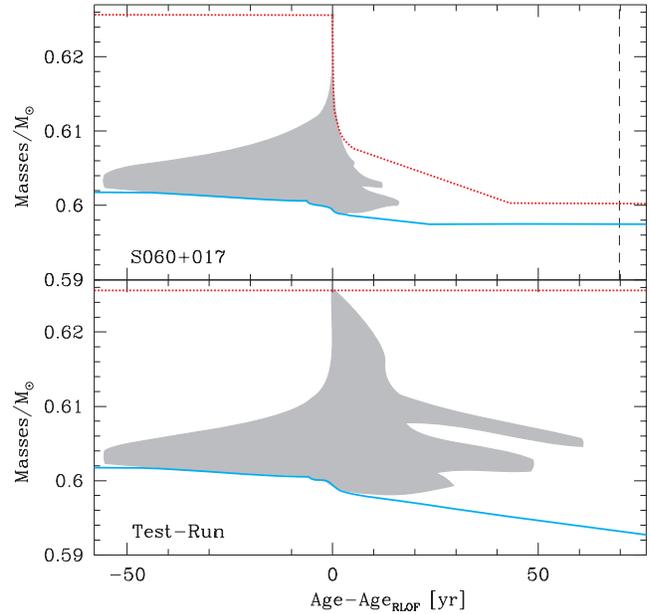}
	\caption{Upper panel: time evolution of the flash driven convective shell (gray zone), of the total mass 
  (dashed line) and of the mass coordinate of the He-burning shell (solid line) for the accretor in the S060+017 system. 
  The vertical dashed line in the upper panel marks the end of the RLOF episode.
  Lower panel: behaviour of the same quantities at the same time in the Test model where RLOF occurs later.
}   
\label{fig06}
\end{figure}

To make this more clear, we performed the second run (Test model) increasing arbitrarily
by a factor of 100 the binary separation at the epoch of the onset of the RLOF overflow in the 
S060+017 system. Note that, as a consequence of such a variation, the Roche lobe of the accretor 
became $\simeq$ 4\rsun, while the Roche lobe radius of the donor increased
to 2.2 \rsun\ and, hence, the mass transfer came to a halt. 
In this way we constructed a model having exactly the same mass of the He-rich buffer of the accretor 
as in the S060+017 system, but which ``freely'' expanded in the space before the onset of the RLOF, 
as in the computations of Paper I.

In the upper panel of Fig.~\ref{fig06} we show the time evolution of the flash-driven convective shell 
(gray area), of the total mass (dotted line) and of mass coordinate of the He-burning shell (solid line) 
for the accretor in S060+017 system.
In the lower panel we show the behaviour of the same quantities, but for the Test model, 
which has larger Roche lobe.
The epoch t=0 corresponds to the onset of the RLOF episode in the S060+017 system, while the dashed 
vertical line marks its end. 
As it is seen, the mass loss triggered by the RLOF episode determines the rapid disappearance of the convective shell. 
As discussed in Paper I, convection affects the evolution of the He-flash in two opposite ways:
on one hand it removes energy from the inner zones of the He-rich buffer and redistributes it 
over the whole envelope, thus reducing the local increase of temperature and damping the thermonuclear 
runaway. 
On the other hand, convective mixing dredges down fresh helium, feeding the burning-shell and, hence, 
powering the flash. 
As a result, the strong and rapid reduction of the convective shell in the S060+017 system limits the 
amount of nuclear energy produced via He-burning and, hence, the resulting 
flash is weaker with respect to the Test model. 
To make this conclusion more quantitative, we define for the model S060+017 two time intervals: 
$\Delta t_1=55.468$\,yr, lasting from the onset of the flash-driven convective shell to the onset 
of the RLOF, and $\Delta t_2=69.774$\,yr, lasting from the onset to the end of the RLOF. In the 
S060+017 model during $\Delta t_1$ He-burning delivers an amount of energy equal to 
$\varepsilon_1\simeq 2.678\times 10^{48}$ erg, while during $\Delta t_2$ it delivers 
$\varepsilon_2\simeq 1.639\times 10^{48}$ erg. In the Test model, the energy delivered 
during $\Delta t_2$ is by a factor 2 larger ($\varepsilon_2^{TM}\simeq 3.219\times 10^{48}$
erg)\footnote{Note that during $\Delta t_2$, the surface radius of the accretor in the S060+017 
system remains practically unaltered, as it is fixed by the Roche lobe radius, while in the Test 
model it increases by a factor 25.}.
\begin{figure}         
 \centering
  \includegraphics[width=\columnwidth]{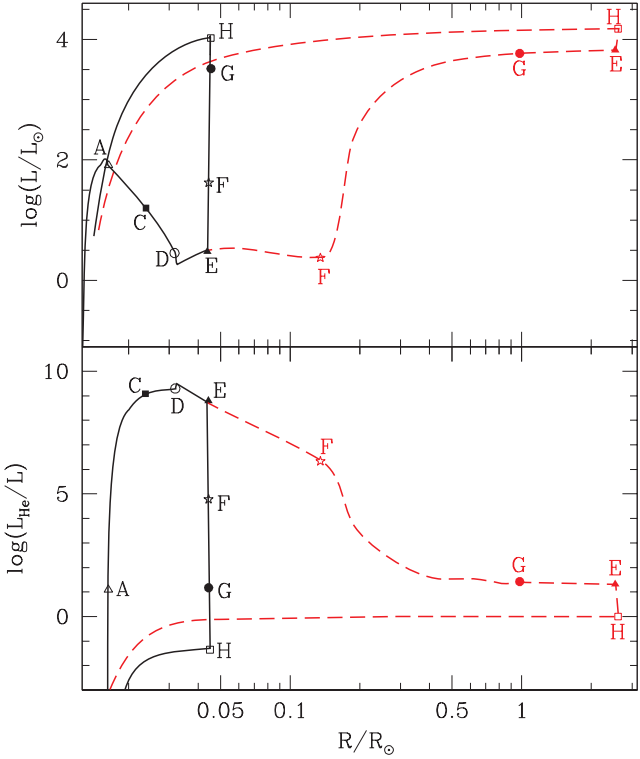}
    \caption{Dependence of the total luminosity $\mathrm{L}$  (upper panel) and ratio of luminosity of the 
             He-burning shell $\mathrm{L_{He}}$ and total luminosity $\mathrm{L}$ (lower panel) on the radius of 
             the CO WD in the sequence S062+017 during the He-flash. S062+017 system is shown with solid lines,
             Test model with dashed ones. In the evolutionary phase where the two accretors evolve 
             identically, dashed lines are hidden behind the solid ones. Different points and letters 
             mark various important epochs, as in Fig. \ref{fig02}.}
\label{fig07}
\end{figure}

Figure~\ref{fig06} also demonstrates that the extension and lifetime of the flash-driven convective shell 
affects the heating of the layers below the He-burning shell. This is clearly indicated by the inward 
shift of the He-burning shell (see solid lines in Fig. \ref{fig06}): 
a larger amount of thermal energy produced via He-burning determines a more efficient heating of the underlying 
zones and, hence, a deeper inward shift of the He-burning shell itself.

To illustrate further the difference between these two cases, i. e., 
between the evolution of very compact systems and that of wider ones,
we present in Fig.~\ref{fig07} the dependence 
of the total luminosity of the accretor $\mathrm{L}$
(upper panel) and the ratio of the He-burning shell luminosity and the total luminosity
$\mathrm{L_{He}/L}$ (lower panel) on the WD radius. The evolutionary paths of two accretors coincide up to RLOF epoch 
in the S060+017 system. In the latter, soon after the onset of the RLOF, convection 
is aborted and the He-rich layers above the He-burning shell, previously heated by the flash, are lost in 
a short time scale (Fig. \ref{fig06}). This almost extinguishes He-burning very rapidly and the only 
energy source to maintain thermal equilibrium becomes contraction. This can be seen in the lower panel
of Fig.~\ref{fig07}, showing that after the RLOF episode in the system S060+017 the 
ratio $\mathrm{L_{He}/L}$ becomes $\simeq$1/20.
As He-burning dies completely, contraction will remain the only energy source balancing the radiative 
losses from the surface since He is never reignited because of the cooling of the He buffer due to the 
interruption of mass-transfer and, later on, to the insufficient heating via accretion (Fig.~\ref{fig05}).
In the Test model, convection continues to support the He-burning shell and, consequently, the delivered 
energy drives the expansion of the WD He-rich envelope to the large Roche radius. 
The dwarf recedes from its Roche lobe when the energy produced via He-burning 
equates the energy losses from the surface. However, this occurs slower than in the case of 
the tight system -- for the same luminosity level after the RLOF, the accretor
in the S062+017 model is more compact then in the Test model (upper panel of Fig.~\ref{fig07}).  
Later on, the two models converge to the same evolutionary path. 
\begin{figure}  
 \centering
  \includegraphics[width=\columnwidth]{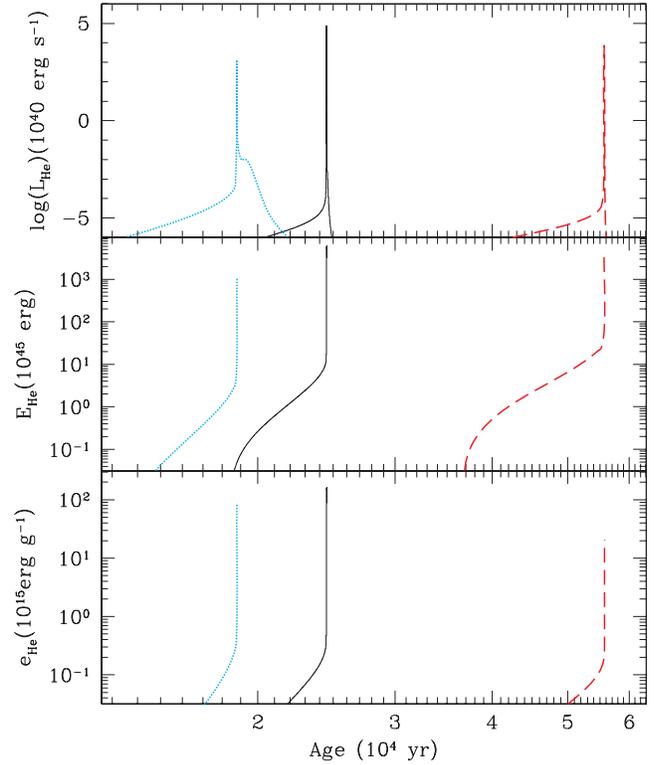}
  \caption{Comparison of the properties of the first flash in the S060+017 (dashed lines), 
           S092+015 (solid lines), and S102+020 (dotted lines) systems. 
           The panels (top to bottom) show the luminosity of the He-burning shell of accreting dwarf, 
           the total energy released during the flash, and the ``specific energy'' of the flash (see text for details).
           The age of the S102+020 system has been arbitrarily increased by $1.9\times 10^{4}$ yr.} 
\label{fig08}
\end{figure}

The difference illustrated above has a sizable effect on the retention efficiency of the 
two models. In particular, in the Test model a larger amount of nuclear energy is delivered 
before the onset of the RLOF episode, so that the the whole He-buffer above the CO core is 
heated more than in the S060+017 system. As a consequence, the He-burning shell can move 
inward; moreover, 
even if a part of the energy injected in the He-rich buffer is dissipated via mechanical work 
(expansion to larger radius), the thermal content of the He-rich mantle remains very large.
As a matter of fact, a larger amount of matter is lost during the RLOF episode in the Test model 
($\mathrm{\Delta M_{lost}=2.7\times 10^{-3}}$\ms), so that the corresponding retention 
efficiency reduces to \acef=0.065 (about 45\% lower than the one obtained for the S060+017 system).
In Paper~I we found that the ``heating'' flash in the M060 {\sl Cool model} erodes 
also part of the pre-existing He-rich buffer, while in the Test model we find a small, but in any 
case positive, retention efficiency. Such a difference reflects the different values of 
\mdot\, adopted for the above mentioned models: in Paper~I we use \mdot=$10^{-7}$\myr, while 
in the Test model the accretion rate is definitively larger: initially it is as high as 
$8.8\times 10^{-7}$\myr\ and, in any case, it remains larger than $3.3\times 10^{-7}$\myr). 
In Paper~I we explored the dependence of \acef\ on the assumed Roche lobe radius, by varying the 
latter in the range 1 -- 45 \rs and we found that the difference in the estimate of retention 
efficiency is smaller than 7-8\%. Such a conclusion is appropriate for binaries with relatively 
massive He-star donors, but, as demonstrated above, not for low-mass compact stars.  

To summarize, in the S060+017 system, due to the small separation, the first He-flash 
does not result in an efficient heating of the layers above the CO core, so that the 
following evolution proceeds substantially different from what derived in Paper I.

\subsection{S092+015 System}
\label{s:s092}

The first flash in the S092+015 system is much stronger than in the other systems, as 
clearly displayed in Fig.~\ref{fig08}, where we plot as a function of the evolutionary time 
three quantities characterising the He-flash for all the systems listed in Table~\ref{tab2}:
the He-burning luminosity $\mathrm{L_{He}}$ (upper panel), the total energy released during the 
flash $\mathrm{E_{He}}$ (middle panel), and the ``specific energy'' of the flash $\mathrm{e_{He}}$ 
(lower panel), defined as the ratio of $\mathrm{E_{He}}$ and mass of the He-rich zone.

The maximum luminosity in the burning shell $\mathrm{L_{He}}$ attained during the 
flash is 10 times larger than for the S060+017 system and about 50 times larger than for 
the S102+020 one (see also Table~\ref{tab2}).
Such an occurrence is not related to the physical properties of the He-shell and to the thermal 
content of the underlying CO core at the onset of the mass transfer, but is determined by the 
adopted combination ($\mathrm{M_{don}, M_{acc}}$) which sets the value of \mdot\, after contact.
At the beginning of the mass transfer, \mdot\, for the S092+015 system is $\simeq 5.0\times 10^{-7}$ 
while for the S060+017 and S102+020 systems it is $8.8\times 10^{-7}$ and $2.2\times 10^{-6}$, 
respectively. As a consequence, 
the compressional heating of the He-buffer occurs at a lower rate so that at the onset of the He-flash 
the He-shell has more degenerate physical conditions (see the values of $\mathrm{\rho_{He}\ and\ T_{He}}$ listed in column B 
of Table~\ref{tab2}).
\begin{table}  
\caption{
Selected physical properties for each of the five strong pulses 
         experienced by the accretor in the S092+015 system. 
         $\mathrm{M_{He}}$ (the mass coordinate of the He-burning shell, \ms),  
         $\mathrm{\Delta M_{He}}$ (the mass above the He-burning shell, $10^{-3}$\ms), 
          $\mathrm{P_{orb}}$ (the orbital period, in min),
         $\mathrm{\rho_{He}\ and\ T_{He}}$ (density in $\mathrm{10^4 g\cdot cm^{-3}}$ and temperature in $10^{8}$ K 
         at the He-burning shell) refer to the epoch of He-ignition.
         $\mathrm{L_{He}^{max}}$ is the maximum luminosity of the He-burning attained 
         during the He-flash. 
         $\mathrm{\left\langle\dot{M}\right \rangle}$ is the mean value of the mass accretion rate 
         (in $10^{-7}$ \myr) during 
         the time lasting from the onset of mass transfer and the onset of the RLOF episode.
         $\mathrm{\Delta M_{tran}}$ and $\mathrm{\Delta M_{lost}}$ (in $10^{-3}$\ms) are the amounts of 
         mass transferred from the donor and lost by the accretor during the RLOF, 
         respectively. \acef\ represents the accumulation efficiency, while $\mathrm{\Delta t}$ 
         (in yr) is the time lasting from the end of the RLOF to the re-onset of mass transfer.
         }
\label{tab3} 
\centering 
  \begin{tabular}{l r r r r r }
   \hline\hline
  & $1^{st}$ & $2^{nd}$ & $3^{rd}$ & $4^{th}$ & $5^{th}$ \\ 
\hline
$\mathrm{M_{He}}$                          & 0.9225 & 0.9180 & 0.9178& 0.9189& 0.9203\\
$\mathrm{\Delta M_{He}}$                   & 7.362 & 5.887 & 6.851 & 8.935 & 14.170 \\ 
$\mathrm{P_{orb}}$                         & 4.830 & 5.069 & 5.331 & 5.753 &  6.588 \\
$\mathrm{\rho_{He}}$                       & 11.66 & 9.22 & 10.78 & 14.00 &  22.13 \\ 
$\mathrm{T_{He}}$                          & 1.354 & 1.385 & 1.367 & 1.289 &  1.205 \\ 
$\mathrm{\log(L_{He}^{max}/L_\odot)}$      & 11.311 & 10.787 & 10.967 & 11.592 & 12.520 \\
$\mathrm{\left\langle\dot{M}\right\rangle}$&  4.178 & 3.087 & 2.413 & 1.748 & 1.015 \\
$\mathrm{\Delta M_{tran}}$                 & 10.230 & 6.865 & 6.838 & 9.799 & 15.802 \\
$\mathrm{\Delta M_{lost}}$                 & 12.787 & 6.107 & 6.584 & 9.191 & 15.085 \\
\acef                                      & -0.250 &  0.110 & 0.037 & 0.062& 0.045 \\
$\mathrm{\Delta t}$                        & 6028.9 & 3434.5& 4805.2 & 8401.0 & 23316.8\\
\hline                  
  \end{tabular}
\end{table}
\begin{figure} 
 \centering
  \includegraphics[width=\columnwidth]{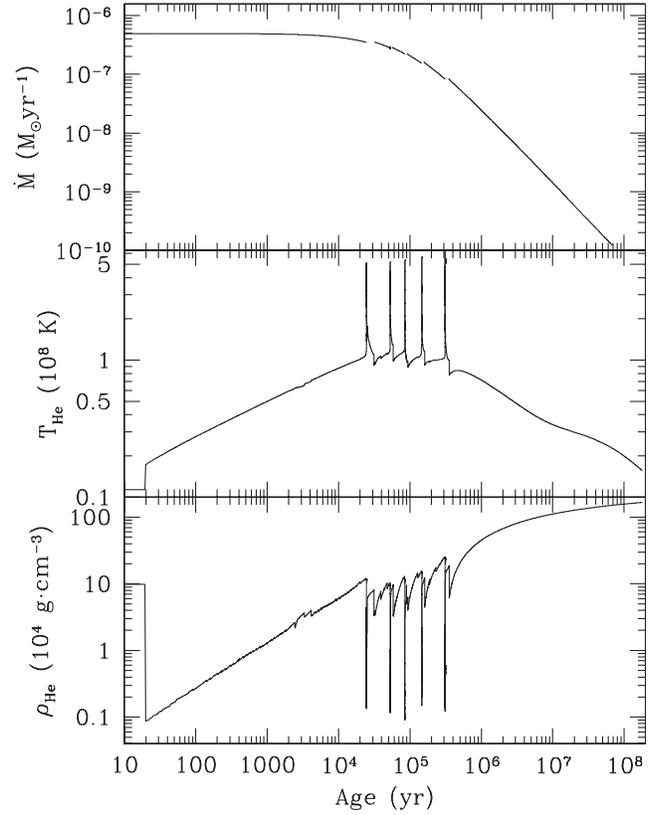}
  \caption{The same as in Fig.~\ref{fig05} but for the S092+015 system.} 
\label{fig09}
\end{figure}
\begin{figure}  
 \centering
  \includegraphics[width=\columnwidth]{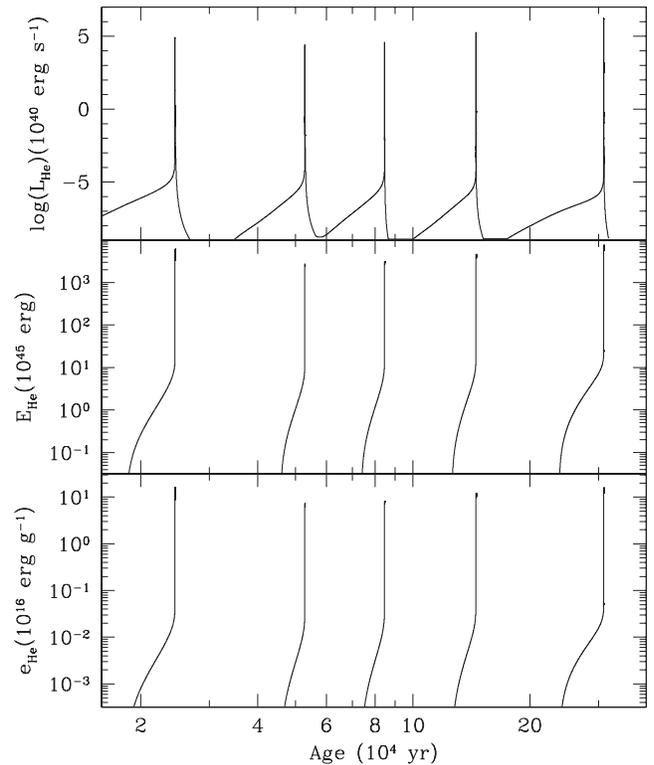}
  \caption{Comparison of the properties of subsequent flashes in the sequence S092+015. The same variables 
           as in Fig.~\ref{fig08} are plotted.}
\label{fig10}
\end{figure}

After the first He-flash, at the re-onset of mass transfer \mdot$\simeq 3.5\times 10^{-7}$\myr;
according to Paper~I, the same initial mass WD, model M092, after the ``heating flash'' experiences 
strong He-flashes for \mdot$\le 5\times 10^{-7}$\myr.
The S092+015 system undergoes other 4 strong non-dynamical He-flashes before a massive degenerate object is 
formed, (see Table~\ref{tab3} and Figs.~\ref{fig09},\ref{fig10}). In Table~\ref{tab3} 
we list selected physical quantities of the accretor during the 5 He-flashes experienced by this system.

By adopting for \mdot\, the mean value before the $2^{nd}$ flash reported in Table~\ref{tab3} and interpolating 
the data reported in Table 4 of Paper I we obtain \acef$\sim 0.61$, definitively larger than the value we 
derive in the present work\footnote{According to the discussion in Sect. 3.1 of Paper I, the evolutionary 
outcome of WDs accreting mass with a time-dependent accretion rate depends mainly on the current value  
of \mdot\, and to a less extent on the previous thermal history. At the onset of the $2^{nd}$ He-flash in the S092+015 system we find 
\mdot$\approx 2.5\times 10^{-7}$\,\myr; for this \mdote, according to the data in Table 4 of Paper~I, 
we obtain \acef $\simeq$0.49.}. As discussed in the case of the S060+017 system, such a discrepancy has to 
be ascribed to the non-efficient heating  of the He-layer during the first He-flash due to the sudden and 
sharp decrease of the mass extension of the flash-driven convective shell. 
When mass transfer resumes after the 5$^{th}$ flash, \mdot$=7.25\times 10^{-8}$\myr\ and the He-shell starts 
to heat up once again. At \mdot$\sim 4\times 10^{-8}$\myr\ the accretor enters the regime which 
for the actual $\mathrm{M_{WD}}$ and constant \mdot\, would correspond to dynamical flashes 
(Paper~I).
In the case under analysis, as \mdot\, continuously decreases, neutrino cooling starts to dominate over the
compressional heating and no additional He-flashes (dynamical or not) are ignited. Hence, accretor turns 
into a nuclearly inert degenerate object which gradually increases its mass.

\subsection{S102+020 System}
\label{s:s102}

The evolution of the S102+020 system after the first flash is quite different. 
Figure~\ref{fig03} and Table~\ref{tab2} reveal that for this system the evolution during the 
first He-flash episode is more similar to that of a ``freely'' expanding model. In fact the flash 
driven convective shell attains the stellar surface \textit{before} the epoch of the maximum 
He-shell burning luminosity and, in addition and more important, the RLOF episode starts
\textit{after} the flash-driven convective zone has receded from the stellar surface (see the sequence 
of the various phases in Fig. \ref{fig03} and the negative value of $\Delta t$ for epochs D and F in 
Table~\ref{tab2}). This means that the He-flash succeeds in heating the whole He-buffer to a higher 
level with respect to what occurs in the S060+017 and S092+015 systems. 
Such an occurrence is a direct consequence of the fact that, when the components in this
binary system come to contact for the first time,
the mass transfer rate is very high (\mdot$\simeq 2.2\times 10^{-6}$\myr), so that gravitational energy 
released by accretion efficiently heats up the entire He-buffer.
This determines less degenerate physical conditions at the onset of the 
He-flash, as compared to the other two considered systems (see $\rho_{He}$ and $T_{He}$ for the epochs 
A and B in Table \ref{tab2}), so that the resulting He-flash is less strong (see the value of $L_{He}$ 
at the epoch C in Table~\ref{tab2} and Fig.~\ref{fig08}). 
\begin{figure} 
 \centering
  \includegraphics[width=\columnwidth]{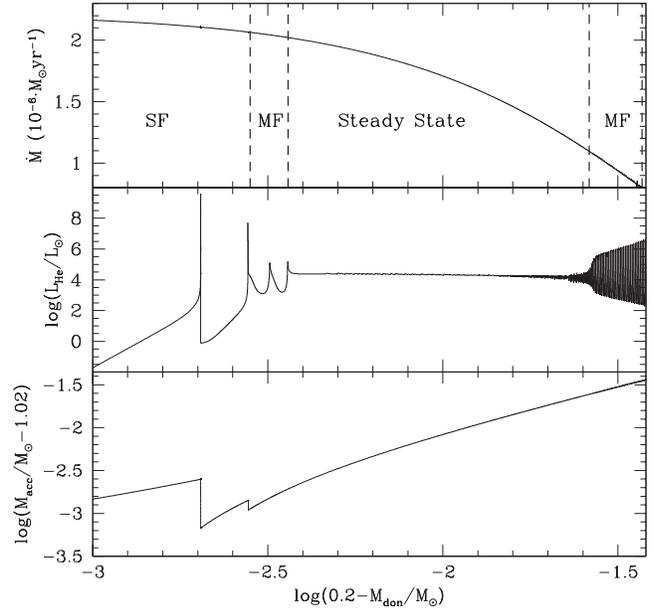}
  \caption{Evolution of the S102+020 system. We report as a function of the mass transferred 
  from the donor the mass transfer rate \mdot\, (upper panel), the luminosity of the He-burning shell 
  (middle panel) and the mass of the accretor $\mathrm{M_{acc}}$ (lower panel). Vertical dashed line in 
  the upper panel marks the transition from one accretion regime to another (see text for details).
}
\label{fig11}
\end{figure}

When mass transfer from the donor resumes, \mdot\, is still as high as $\sim 2.1\times 10^{-6}$\myr 
(see upper panel in Fig. \ref{fig11}). In Paper~I we found that the ``heated'' M102 model accreting at 
such a rate experiences quiescent burning of He. 
On the contrary, in the present work the accretor in the S102+020 system experiences other 3 flashes. 
The flashes become progressively less strong (see middle panel of Fig.~\ref{fig11}) and the corresponding 
retention efficiency increases (see lower panel of the Fig.~\ref{fig11}). 
In particular, we find \acef=0.562 for the second flash and \acef=1.0 for the third one. 
We classify the latter as a ``mild flash'' (MF), i. e., a flash which releases so little nuclear 
energy that the accretor remains confined well inside its Roche lobe and no mass-loss occurs. 
Such a behaviour is determined by the fact that, pulse by pulse, the He-rich buffer and the 
most external layers of the underlying CO core heat up, thus attaining 
the physical conditions suitable for quiescent He-burning.
The resulting evolution is completely 
different from that of the accretor in the S092+015 system and it is a direct consequence 
of the mass transfer rate after the first flash episode which  depends 
on the parameters of the initial binaries. 
In particular, as already mentioned before, \mdot\ in the S092+015 system 
after the first flash should determine recurrent strong flashes also in a fully heated  model 
with the same total mass and mass of the He-buffer (e. g. the ``heated'' M092 model 
in Paper~I). At variance, in the S102+020 system the mass transfer rate is so high that 
the released gravitational energy  prevents the cooling down of the 
He-burning shell during each inter-flash period.
As displayed in Fig. \ref{fig11}, after four flashes the accretor reaches Steady burning regime 
(Steady State), which lasts as long as, due to continuous decrease of the mass transfer rate 
from the donor, the extension of the He-buffer above the He-burning shell reduces below a critical 
value and the accretor enters again the mild flashes regime. 
The transition from one regime to 
another occurs smoothly so that we arbitrarily define the 
value of \mdot\, at which such a transition occurs as the epoch when the maximum luminosity of the 
He-burning shell during the He-flash becomes twice the value of the surface luminosity along the 
high luminosity branch. Under this assumption we find \mdot(SS-MF)$\simeq 1.09\times 10^{-6}$\myr.
Due to the evolution of binary parameters, the mass transfer rate continuously decreases and when 
it becomes lower than $\sim 8.3\times 10^{-7}$\myr, after 29 MFs, 
Strong Flashes start again. In the upper panel of Fig. \ref{fig11} the transitions from one regime to another 
are marked by dashed vertical lines. 

\begin{figure}   
 \centering
  \includegraphics[width=\columnwidth]{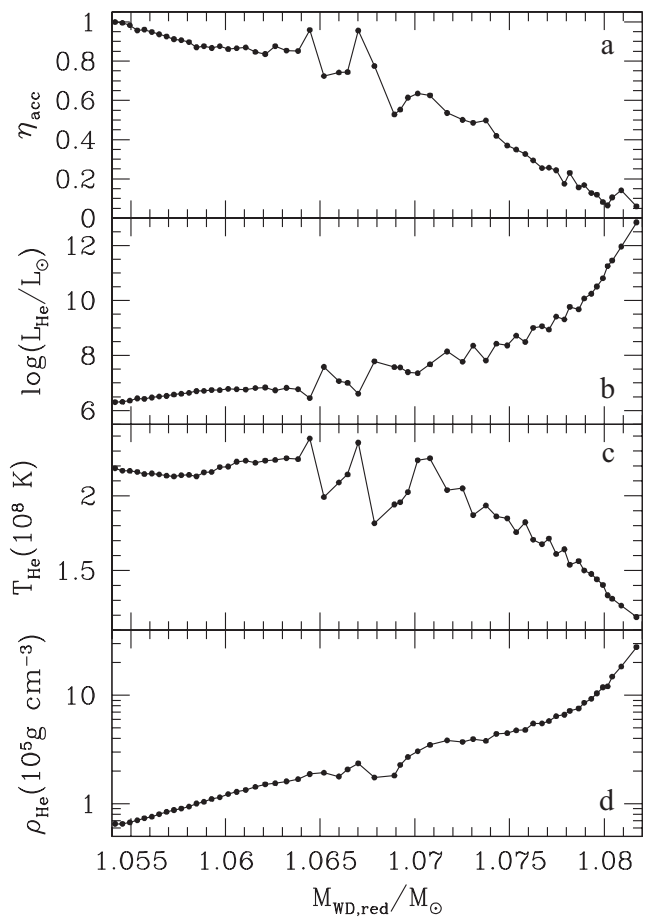}
  \caption{Evolution of the accretor in the S102+020 system during the Strong Flashes accretion regime. 
  We report, as a function of the total mass of the accretor at the bluest point along the loop in the HR
  diagram ($\rm{M_{WD,blue}}$) the value of the retention efficiency (panel a), the luminosity of the 
  He-burning shell (panel b) and the temperature (panel c) and density (panel d) at the epoch of He-ignition.
}
\label{fig12}
\end{figure}

The following evolution is described in Fig. \ref{fig12} where we show, as a function of the accretor 
total mass at the bluest point along the loop, the value of the retention efficiency,
 and the physical 
conditions (temperature and density) at the epoch of the He-flashes ignition. 
As well, the maximum luminosity in the He-burning shell $\mathrm{L_{He}}$ is shown.
As expected for a monotonically decreasing mass transfer rate, \acefe, on average, decreases continuously 
(upper panel), as the physical conditions at the He-shell become more degenerate and, hence, the strength 
of successive flashes increases. 
The general trend is defined by two main factors.
On one hand, as the accretion rate decreases and \acef\ reduces, the time span after the end of the 
RLOF episode and the mass transfer resumption increases, so that the the He-buffer cools 
more. This determines the increase of the amount of mass to be transferred to ignite the 
successive He-flash which, as a result, is also stronger. 
On the other hand, pulse by pulse, the Roche lobe radius of the accretor increases, so that 
during the successive He-flashes the heating of the whole He-buffer is more efficient, as the RLOF occurs 
when a larger amount of nuclear energy has been delivered (see above the 
 discussion of S060+017 system and related Fig.~\ref{fig06}). 
Though, the evolutionary curves also exhibit irregularities whose 
origin depends on the complex interplay of various additional factors.

We find that after the $58^{th}$ strong flash, 
the mass of the donor has reduced to $\sim$0.0875\msun\ and that of the accretor  increased
to $\sim$ 1.0822\msun. 
When accretion resumes, after 181916 yr, the mass transfer rate is $4.2\times 10^{-8}$\myr\ and it 
rapidly decreases so that the compressional heating produced by accretion is 
not able to balance the radiative and neutrino cooling of the whole He-rich buffer. Hence, as already 
illustrated for the S060+017 and S092+015 systems, the accretor in the S102+020 system cools down and 
the final outcome is a massive CO core surrounded by an extended He-buffer.

\subsection{``Cold'' S060+017 Model}
\label{s:cold}
In order to investigate the dependence of our results on the cooling age of the accretor,
we let the ``{\sl heated model''} M060 from Paper~I to cool for 2 Gyr which, according to population 
synthesis computations, can be considered as the typical time of formation of an \am\ system from a pair 
of detached WDs \citep{ty96}. 
After that, we start the accretion following the same procedure as above. The results are 
summarised in Table~\ref{tab2}.
As it can be noticed, at the beginning of the mass transfer, the physical conditions at the base of the 
He-burning shell are more degenerate (temperature is a factor 5 lower, while density is practically 
the same). Notwithstanding, the amount of mass to be transferred to ignite the He-flash 
is only 3\% larger and the ignition conditions are very similar (compare columns A and B in Tabs. \ref{tab2}
for the standard and ``Cold'' S060+017 Models) and the resulting He-flash has almost the same strength. 
During the RLOF episode the cold S060+017 system loses about 11\% more mass so that the final retention 
efficiency is lower. As a consequence, the post-RLOF episode separation is a bit larger and, hence, the 
time span up to the re-onset of mass transfer increases by more than a factor 3.5. 
In any case, the evolution after the He-flash episode occurs exactly as in the S060+020 system: the mass 
transfer rate decreases very rapidly, so that the radiative and neutrino losses from the He-buffer become 
dominant and the system evolves to the formation of a massive CO WD with an extended He-rich mantle.

The results for the ``Cold'' model suggest that the models we considered are
representative also for systems  with longer cooling age. The reason is that
at the luminosity level 0.01 \ls\, the physical conditions at the center of WD and at the base of 
the He-rich buffer have attained an asymptotic value. 
Increasing the cooling age of accreting WDs from 800\,Myr to (2 -- 3)\,Gyr practically does not 
affect their reaction to accretion\footnote{Longer formation times are not considered in 
the present analysis as they are not typical for \am\ stars.}.
Such a statement is also confirmed by the computation of the evolution of the ``Cold'' 
analog of S092+015 system up to the onset of the first RLOF episode.
The released nuclear 
energy was quite similar to the one in the ``hot'' model,  while the accreted mass was only 1.05\% larger.
This similarity suggested us to terminate this additional extremely time-consuming computation, 
since we infer that the initial temperature of WD will not play any role, like in the case
S060+017.

\section{Discussion and Conclusions}
\label{s:diss}

In the present work we studied accretion from He WD donors onto CO 
WDs in ultracompact \am\ binaries (IDD). At variance with our earlier study of the He-burning regimes 
in WDs accreting mass at constant \mdot\, (Paper~I), in our current analysis we adopted time-dependent 
accretion rates, as determined by the loss of angular momentum via GWR. 
As in Paper~I, we assumed that, if He-flashes on the accretors result in RLOF, matter is 
lost from the system until accretor recedes from the critical lobe. The matter leaving the 
system has the specific orbital angular momentum of the accretor. The systems considered in our analysis 
have the following properties: (i) mass transfer rate permanently declines and (ii) accretion may be 
interrupted due to mass and angular momentum losses. In Paper~I we assumed that the accretion process 
is almost continuous, as it resumes soon after the end of the RLOF episode. 
At the end, in our earlier study we preset arbitrarily the accretor Roche lobe radius, 
while in the present work the latter is defined by the actual parameters of the given binary system, 
i. e., masses of the components. The initial values of them were taken corresponding to
supposed precursors of \am\, stars --- extremely low white dwarfs systems (\elm s). The donors were 
approximated as pure He-objects with zero-temperature, proper to the formation timescale of \am\, stars 
$\sim$Gyr.

The actual  He-burning regime onto the accretors in IDD is defined by 
the interplay of several factors. 
The first, as in other accreting systems, is the balance between the compressional heating driven by 
the mass deposition and the cooling of the He-rich buffer via inward thermal diffusion and neutrino 
emission. The second is the permanently decreasing accretion rate.
The third is the degeneracy level of the physical base of He-rich buffer. 
Last but not least, the Roche lobe radius of accretors may be very small, down to several 0.01\,\rsun. 

As expected (see the discussion in Paper~I), we found that after the onset of mass transfer onto the accretors, 
the He-burning shell above the CO core heats up and, when the nuclear timescale for 3$\alpha$-reactions
becomes shorter than the local thermal adjustment timescale, a nuclear flash occurs. Due to the degenerate 
physical conditions, this first flash is strong and it drives to a significant expansion of the star. 
It is worth noticing that this first flash is the equivalent in the real world 
of the ``artificial'' initial  {\sl heating} flash used in the studies of He-burning onto WDs, 
in order to mimic post-AGB objects. However, in a typical \am\ system the accretor 
\textit{should} be cold, since the timescale of \am\ stars  formation is $\sim$\,Gyr.
As a matter of fact, such a flash results in a RLOF by the accretor 
and drives to the loss of mass and angular momentum from the system. 

As shown in Sect.~\ref{s:s060}, at variance to ``wide'' interacting systems, 
the post-flash evolution of IDD is affected by the prompt mass loss which limits the lifetime 
and extension of the flash-driven convective shell, thus limiting the feeding of the He-burning shell by 
convection and reducing the total amount of nuclear energy released during the flash. 
Such an occurrence prevents the efficient heating of the pre-existing He-rich layers below the 
He-burning shell and causes the rapid extinction of nuclear burning. Hence, in the 
post-RLOF phase the accretors in IDD contract very rapidly in order to maintain the 
thermal equilibrium.

According to the accepted paradigm for the evolution of outbursting binaries, the matter overflowing 
the accretor Roche lobe leaves the system, taking away the specific angular momentum of the 
accretor. As a result, the system becomes detached, mass transfer comes to a halt and the WD cools. 
Hence, the possibility of resumption of} nuclear activity depends on the duration of this ``hibernation'' 
phase and the masses of the two components which, in turn, determines the value of \mdot\ at the 
the resumption of mass transfer. As it has been found for the sequence S062+017 (Sect.~\ref{s:s060}), 
for an ever decreasing \mdot, it is possible that the release of gravitational energy occurs in a 
so long timescale that it can not counterbalance the cooling by inward thermal diffusion and neutrino 
emission and another flash can not be ignited.
For the other systems considered in the present work, this limits the total number of flashes 
experienced by a given system.  

The rough comparison of the present study and Paper~I suggests that the expected number of nuclear flashes 
in IDD is a factor $\sim2$ lower than the one expected by interpolating the results based on 
models with constant \dotm. This reduces further the estimation obtained in Paper~I of only several 
thousand ``nuclearly active'' IDD currently present in the Galaxy.

The results obtained in the current work exhibit a very mild dependence on the initial thermal 
content of the accretor, for cooling ages in the range 0.8 -- 2 Gyr (see Sect.~\ref{s:cold}), 
so that they can be considered as representative of the entire population of IDD. 

For each fixed initial CO WD mass, as the mass transfer rate is permanently decreasing, the last 
flash experienced by accreting WDs is the strongest one (see Figs.~\ref{fig09}, \ref{fig10}, 
\ref{fig12}). As mentioned before, \citet{2007ApJ...662L..95B} suggested that in a series of flashes 
the last one may be of dynamical nature (\snia). None of the systems considered in the present work develops 
the physical conditions suitable for such a dynamical flash. In particular, we found that 
the mass of the He-buffer at the onset of the last He-flash is always lower than the critical value 
necessary to obtain a \snia, derived by Bildsten and his coauthors for the same 
WD mass (see Fig.~2 in \citeauthor{2007ApJ...662L..95B} \citeyear{2007ApJ...662L..95B} and Fig.~1 
in \citeauthor{2009ApJ...699.1365S} \citeyear{2009ApJ...699.1365S}). Such a difference has to be ascribed 
to the different thermal content of the layer below the He-burning shell at the onset of the last He-flash. 
In our model, the previous accretion history determines the partial heating of the most external zone of the CO 
WD and of the 
He-burning shell, so that at the beginning of the last mass transfer episode driving to a He-flash 
we have $T_{He}\sim 10^8$ K, for both S092+015 and S102+020 systems. 
As a consequence the ``last flash'' in our computations will never attain physical conditions 
suitable to synthesize iron peak elements. 
Thus all three considered systems have 
the same fate: transformation of the accretor into a massive degenerate object with a CO-core, which has almost the 
same mass as the initial accretor, and a massive He-buffer, representing a large part of the 
initial donor. 

Recently, \citet{brooks2015} performed an analysis similar to our one, but focused on \am\, systems with a 
non-degenerate He-burning star as a donor.  
Their results can not be compared directly with our findings, 
because the response to mass extraction of a He-star is different from that 
of a degenerate object and, hence, the resulting evolution of the host binary system has to be 
different. \citet{brooks2015} found that the last He-flash experienced by the accretor 
of the ``He-star family'' of \am\, systems is never strong enough to produce a \snia\, event. So, by combining 
their results with our one, we can conclude that \am\, family at large will never produce an explosive event 
at the end of their life. 
We remark also, that  we follow in detail the thermal evolution of the accretors during each flash episode, 
determining the effective accumulation efficiency, while \citet{brooks2015} remove the entire He-envelope 
above the He-burning shell at the onset of the thermonuclear runaway.

\citet{brooks2015} claim that the first He-flash experienced by \am\ systems with non-degenerate donors more 
massive than 0.4\msun\ and accretors more massive than 0.8\msun\ is ``vigorous enough to trigger 
a detonation in  the helium layer'', which could produce either a \snia\, or a real 
type Ia Supernova, if also the CO core detonates. Interestingly enough, this first explosive event could destroy the host binary, 
thus reducing the expected number of \am\, systems from this evolutionary channel. This could help in solving the 
problem of the apparent deficiency of observed \am\ stars \citep{2013MNRAS.429.2143C}.
At variance, we find that the first He-flash occurring  
in the IDD explored in the present study never develops the physical conditions suitable for a detonation in the 
accreted He-rich buffer. This difference reflects the different behaviour of the mass transfer rate before 
the onset of the He-flash: in particular, for IDD \mdot\, is continuously decreasing in time, 
while it is rising in the binaries studied by \citet{brooks2015}. 
In this regard, it is worth noticing that \citet{brooks2015} consider \am\, systems 
with He-stars unevolved or only slightly evolved at the onset of the RLOF. However, 
the mass transfer history from He-stars depends on the extent 
of the donor evolution prior to the RLOF. As a matter of fact the \mdot\ curve may be rising, 
almost flat or very slightly decreasing \citep{yungelson2008}. Moreover, material transferred from evolved He-star donors 
has lower He abundance reducing the probability of strong outbursts.
\begin{figure}    
 \centering
  \includegraphics[width=\columnwidth]{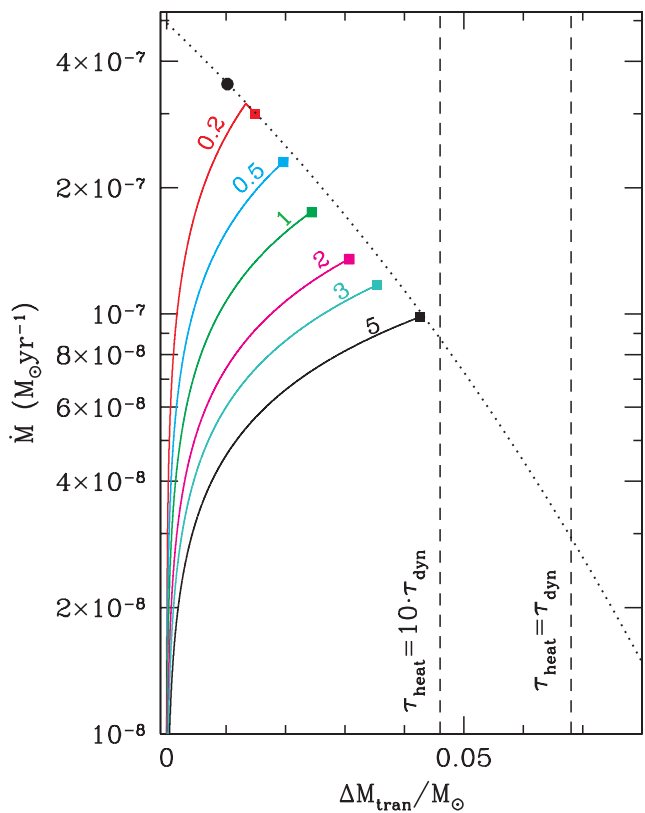}
  \caption{Mass transfer rate in the S092+015 system as a function of the mass transferred from the donor $\Delta M_{tran}$ 
           in different timescales $\tau$ (in Myr), as labelled (solid lines). 
           The dashed line represents accretion rate $\dot{M}_{FD}$ 
           corresponding to fully conservative mass transfer. Filled circle marks the coordinate 
           ($\Delta M_{tran},\dot{M}$) 
           at the onset of the thermonuclear runaway for the S092+015 system under ``standard''  assumptions. 
           Filled squares represent the same quantities for 
           the models computed with different $\tau$. The two vertical lines mark the critical mass of the He-buffer for a 
           0.92\msun\ CO WD corresponding to the condition $\tau_{heat}=10\tau_{dyn}$ (left) and $\tau_{heat}=\tau_{dyn}$ (right) 
           as derived from \citet{brooks2015}.
           }
\label{fig13}
\end{figure}

We recall that in our computation we model the He-WD donor as a zero-temperature object, i. e. as a completely 
degenerate one. In the real world, however, the most external layers of WDs, independently of their chemical 
composition, are only partially degenerate or not degenerate at all. As a consequence, when they fill 
their Roche lobes, the mass transfer rate increases in time, as it occurs for
homogeneous He-star. Once these surface layers have been lost, 
and the fully degenerate inner region starts to be removed, \mdot\, decreases \citep[][see also Fig.~13 in Paper~I]{2007MNRAS.381..525D}. 
The duration of the first phase of mass transfer depends mainly on the mass of the non-degenerate layers and, in turn, 
on the cooling age of the donor. On a general ground, such an occurrence implies that the compressional heating of the accretor should 
occur, at least at the beginning, in a longer timescale, so that the resulting He-buffer at the onset of the He-flash should be larger. 
In order to evaluate quantitatively the effect of this rising phase of the mass transfer rate, we computed several additional models for 
the S092+015 system, by adopting as mass accretion rate:
\begin{equation}
\dot{M}=\min\left(\dot{M}_{FD},{\dot{M}_{FD}^{0}}\cdot\log\left(1-{\frac{t}{\tau}}\right)^{-1}\right),
\label{e:mt}
\end{equation}
where $\dot{M}_{FD}$ is the rate derived with the $M-R$
approximation described in Sect.~\ref{s:model}. 
In the second term, the mass of the partially degenerate/not degenerate layers $\Delta M_{ND}$ is parameterized by 
means of the timescale $\tau$: large values of $\tau$ correspond to large $\Delta M_{ND}$. 
$\dot{M}_{FD}^{0}\simeq 4.978\times 10^{-7}$ 
\myr\ is the value of $\dot{M}_{FD}$ at the onset of the mass transfer.
In Fig.~\ref{fig13} we report as a function of the mass transferred from the donor
the evolution of the mass transfer rate described by Eq.~(\ref{e:mt}), for different 
values of $\tau$. 
This figure discloses that for realistic values of $\Delta M_{ND}\le 0.03$\,\msun, corresponding to $\tau\le$\,2 Myr, 
$ \Delta M_{tran}$ increases up to a factor of $\sim$3, so
that the resulting He-flash is stronger than the one in the S092+015. 
In any case, we do not expect that this could significantly alter the following 
evolution, because, as discussed in Sect.~\ref{s:s060}, the prompt occurrence of
 the RLOF limits the heating of the He-buffer and, hence, the effects on 
the thermal properties of the accretor. Values of $\Delta M_{ND}\ge$0.05\msune, 
corresponding to $\tau >$ 5 Myr, are rather unlikely. 
According to the results of these numerical experiments, we can conclude that a 
dynamical event could hardly arise in IDD.

Helium WDs with masses in the range explored in the present study have non-degenerate outer hydrogen layers
with mass $\le 0.01$\,\msun\ at an age of $\sim$\,1Gyr \citep{2007MNRAS.382..779P}, which is usually considered as the typical 
formation time of \am\ stars. Transfer of this matter onto the companion upon RLOF may cause explosive phenomena
similar to Classical Novae, like in ordinary cataclysmic variables.
Recently \citet{2015arXiv150205052S} suggested that the ejected matter forms a common envelope in which
the components may merge. However, to drive a firm conclusion, it is necessary to consider the interaction of the ejecta with 
the two components, having in mind, of course, the compactness of the system. 
This problem, perhaps, demands 3-D hydrodynamical simulations.  
       
Strong flashes occurring in the \am\ systems considered in the 
present study almost definitely can not be identified with unique Helium Nova V445~Pup 
\citep{ashok03b,ashok03a}, since pre-outburst luminosity of the latter
$\log(L/\ls) = 4.34 \pm 0.36$ \citep{2009ApJ...706..738W} is too 
high for pre-flash CO WD in IDD. Rather, its progenitor may be a massive ($\sim1\,\ms$)
WD accreting at a rate $\sim 10^{-6}$\,\myr\ from a massive (also $\sim1\,\ms$)
He-star companion \citep{2009ApJ...706..738W,2014MNRAS.445.3239P}. A possible
progenitor may be similar to the unique sdO+WD system  
HD~49798 \citep{2009Sci...325.1222M} with both unusually massive components.

\section*{Acknowledgements}
The authors acknowledge useful discussions with G.~Nelemans and M.~Dan.
We acknowledge an anonymous referee for suggestions which helped us 
to improve the presentation of our results.
LP acknowledges support from the PRIN-INAF 2011 project 
``Multiple populations in Globular Clusters: their role in the
Galaxy assembly''. 
AT acknowledges support from the PRIN-MIUR 2010-2011 project 
``The Obscure Universe and the Cosmic Evolution of Barions''. 
LRY acknowledges support by RFBR grants 14-02-00604, 15-02-04053 and Presidium 
of RAS program P-41. \\
This research has made use of NASA's Astrophysics Data System.

\bibliographystyle{mn2e}	
\bibliography{piersanti}
\label{lastpage}

\end{document}